\documentclass[a4paper]{article}
\usepackage{graphicx}
\usepackage{amsmath}
\usepackage{hyperref}
\usepackage{color}

\begin{document}

\begin{center}

{\bf \Large Sociophysics - an astriding science}\\[5mm]

{\large  Krzysztof Ku{\l}akowski$^+$ and Maria Nawojczyk$^*$ }\\[3mm]

{\em

$^+$Faculty of Physics and Applied Computer Science, $^*$Faculty of Applied Social Sciences, 
AGH University of Science and Technology, al. Mickiewicza 30, PL-30059 Krak\'ow, Poland

}

\bigskip

$^*${\tt kulakowski@novell.ftj.agh.edu.pl}

\bigskip

\today

\end{center}










Many intriguing puzzles of today is due to our lack of knowledge on the human 
behaviour. We are surprised how voters in a well-developed country can 
be cheated and put to war by their president and appoint him again to the 
position; how another society brought up in the fine words of love of freedom 
rolls from the eve of democracy down to authoritarianism; how an experienced 
politician deceived by the propaganda of his own advisers gets into a self-destruction
\cite {bbcb,biny,eco}. There is a continuous need to discover laws in social 
world. However, even the tentative status of this target remains unclear
\cite{bro,kind,wei}. A fundamental disagreement lies at the heart of social science 
about whether social phenomena can be subject to the same explanatory goals as physical 
phenomena \cite{a}. Positivist, namely Emile Durkheim \cite{b}, claimed that the 
methods used to study the social world did not differ in any important way from 
the methods used to study the physical world. However, Max Weber \cite{c} offered 
a serious empirical alternative to positivism in the form of hermeneutics that 
held the view there were crucial differences between the physical and social worlds. 
Thus, in our discussion we put forward the questions. Can social research share, with 
physical sciences, the goals of prediction and explanation?  Are the social laws 
different in nature from those of natural
sciences? How universal they should be? Do they allow for any predictions? \\

In these discussions, physics is quoted sometimes as a reference point \cite{rkm}. Moreover, 
physicists are recently more and more involved in research on social
systems - the so-called sociophysics is a new branch of interdisciplinary studies \cite{cc,stf}.
As it is well known, the sociology started from the 'social physics' and 'law of 
three stages' \cite{oxf}. Social research is a child of the scientific age. As an 
investigative discipline, its origins are to be found in nineteenth century model of 
physical science. Thus, empiricism may be defined as the idea that all knowledge has 
its origins in experience that is derived through the senses. However, observation 
is not a straightforward affair for it contains two dimensions which interact in 
complex ways. They are the cognitive and social dimensions. Therefore, there is a 
constant relationship between theory and data. Today the set of methods applied to 
investigate social systems cover a large area between philosophy, anthropology, economy, 
history, linguistics, psychology and literature. Is there still any place for 
physics? In this text we show that once physics is applied to a system described 
with many variables, the concept of law becomes fuzzy. Our example 
is the second law of thermodynamics, which has the long-lasting special status in 
physics.  \\

If any branch of physics can be applied to social systems, the statistical mechanics 
seems to be first in the queue. It has a special status among other branches of physics, 
because it deals with systems
with as many as $10^{23}$ variables. It is obviously impossible to tackle 
so many data; then we limit the description to two or three macroscopically measurable 
parameters, the others are averaged out. The cost of this step is that the system 
dynamics reveals new properties, which were absent in microscopic theories. 
Namely, for each macroscopic state $A$ there is some number $N(A)$ of microscopic states,
which are qualified as $A$ during the macroscopic measurement. As we know, the entropy 
of the state $A$ is defined as proportional to $\ln(N(A))$. The problem is that 
more than often we can distinguish 'disordered'  and 'ordered' states, say $A$ and $B$, 
where $N(A)$ is dramatically
larger than $N(B)$. Let us, then, prepare the system in an initial state $B$ - what 
will happen? The system spontaneously evolves from $B$ to $A$ and will not return 
in time period shorter than, say, the age of the Universe. This is the core of the
irreversibility, of the entropy increase and of the second law of thermodynamics, 
the subjects around which the discussion is still amusing \cite{daw}. But as we see, 
the effect of irreversibility appears because our macroscopic measurements are 
inaccurate. Every microscopic state $\alpha$ is unique and then, $N(\alpha)=1$. Then, 
the second law of thermodynamics is a consequence of our way of doing measurements. 
Most philosophers of science have argued that the method used is the only guarantee 
that the knowledge obtained is scientific \cite{d}. In this sense, science is method. 
Still some questions remain.  Is it objective? Is it a true inherent property of the 
Nature? or just a consequence 
of our theoretical description?\\

In a purely macroscopic theory, i.e. the phenomenological thermodynamics, we ignore 
not only the information on the values of the internal variables, but also their 
mere existence. An example of the state $A$
is the state of a metallic rod, both ends of the same temperature. This state happens 
as a natural consequence of a specially prepared state $B$, when one end is heated and 
another end is yet cold. The process of heat flow is irreversible, macroscopically measurable 
and real. If we are able to follow the time evolution of the microscopic variables, i.e. 
the coordinates and velocities of atoms in the rod, the trajectory of the system would
drive it from one microscopic state to another. In this case there is no need to introduce 
entropy; if one does so, he finds that it is equal to zero and it does not vary in time. The second law of thermodynamic is then entangled with our axioms.\\

Maybe we should add that the second law of thermodynamics is at the root of our understanding
of physics. As it was sarcastically formulated by Arthur Eddington, the famous astrophysicist: 
{\it If someone points out to you that your pet theory of the universe is in disagreement with 
Maxwell's equations  then so much the worse for Maxwell's equations. If it is found to 
be contradicted by observation  well, these experimentalists do bungle things sometimes. 
But if your theory is found to be against the second law of thermodynamics I can give you 
no hope; there is nothing for it but to collapse in deepest humiliation.} \cite{daw}. 
This example of the second law of thermodynamics teaches us that a theory is accepted if 
it reflects the results of our measurements. As such, it depends on the measurement method. 
By employing the correct method, the scientist may be sure that their findings are true, 
repeatable and generalizable. These logical attempts to build a justificatory framework 
for scientific knowledge were ruined in social sciences by Karl Popper \cite{e}. According 
to him a scientific theory, as opposed to a pseudoscientific theory, is one open to falsification. 
If the theory passes the tests, it is not confirmed, it only meant that it was not falsified 
on that occasion. By eliminating untruth through the falsification process, science moves closer 
to the truth. The question if the theory is true drives us too far.\\

The same law, although so atypical in physics, provides a bridge to other sciences. 
Let us take into account that infinite sources of heat do not exist: every one will 
run out after some time. More engines work at the source, more active is the heat flow, 
sooner the source gets cold; 
when it is colder, the efficiency of engines working between the source and the environment 
decreases. In describing an engine as inefficient the engineer has recourse to an agreed specification 
of efficiency for a given engine. If a social scientist described the workforce of a factory as inefficient, 
it is unlikely that there would be agreement over such characterization. Karl Poppers definition of 
facts may be one way around this difficulty. According to him, facts are something like a common 
product of language and reality \cite{f}. What we see as factual is a product of our theories and 
an agreement between them and reality itself. Back to efficiency, similar rule is known in 
economy as the law of diminishing returns \cite{wayne}. 
This argument works without a reference to individual engines or individual enterprises.
Obviously, we can use it without making attempts to 'understand' the souls of the businessmen.
From the microscopic point of view the state when there is something to gain in the market
is rare; in an ideal market as in the equilibrium state there is nothing to gain - there is no 
heat flow. Once however a source of income appears, more and more agents use it and fortwith 
the source is empty. The same law seems to apply to any other systems, including those 
most complex, as the society. As there is much more disordered than ordered states, the 
disorder increases, regardless of the definition of disorder itself. \\

The question by far more important is if the second law will be qualified as useful by the 
sociologists. There are at least two reasons for the negative answer. First is the assumption 
of the second law, which states that the system should be isolated. In economical and social 
sciences, this assumption always remains disputable. Second difficulty is that in a social 
system there are many recognized and unrecognized processes which run with different velocities and 
therefore one could argue that the state of equilibrium is never attained. At the end we have to notice
that in the physical sciences, unlike in social sciences, there is a desire for something stronger than 
predictions that are quite likely to be accurate. Science requires invariable laws of nature in 
order that our predictions about the escape velocity of space shuttles do not end in disaster. 
To the contrary we can find many social scientist who are of the opinion that social researchers 
are not in business of predicting or explaining at all \cite{g}. Because not only is the social 
world more complex than physical world, but it is of completely different nature.\\

Still, the argument on the entropy increase has at least one ready application to the problem 
well established in sociology: the racial segregation problem as discussed by Thomas Schelling
\cite{schl}. There is only one state with perfect order, where two communities live in two separated
areas. Once we allow for some disorder, where the communities mix at least slightly, the number
of possible states abruptly increases. In other words, the states with spaial mixing of the 
communities are much more probable, than the states where the communities are strictly separated. 
This is the origin of a diffusion of members of one community 
to the area of the other and vice versa. Actually, the same mechanism operates at the n-p junction.
Recent computer simulations of the problem set by Schelling were thoroughly described in 
Ref. \cite{stso}. In physics, the quantity which controls the amount of disorder is temperature.
However, as it was pointed out in \cite{stso}, disorder is absent in the Schelling's formulation. 
Once temperature is zero, the tendency to order must prevail. After Schelling himself in 
2005 \cite{harf}: {\it A very small preference not to have too many people unlike you in the 
neighbourhood, or even merely a preference for some people like you in the neighbourhood could lead 
to such very drastic equilibrium results that looked very much like extreme separation.}
In physics, it is only at states of equilibrium where temperature makes sense. Again, there is 
no direct equivalent of temperature in sociology. The authors of Ref. \cite{stso} refer to two
indirect ones: 'tolerance' as indifference and 'trouble' as external conditions other than 
the preference of this or that community. \\

In 2007, Dietrich Stauffer gave a talk in Cracow about the Schelling model \cite{schl, stso} for a 
humanistically oriented audience. Having finished, he got a question: 'What can you say about motivations of 
those people who move from their flats to change neighbors?' The answer was: 'Nothing'. The 
problem of communication between physics and sociology is that the above thermodynamic formulation
does not provide any new perspective for a social scientist.  For a sociophysicist, the keywords for the 
race segregation problem are lattice theory and diffusion \cite{stso}, with temperature and entropy 
as the background. For a sociologist, the keywords could be tolerance, status and prejudice.
As it was stated by Robert K. Merton, a sociological theory is a logically interconnected set 
of propositions from which empirical regularities can be derived \cite{rkm}. He also pointed out 
that there are both internal and external accounts of science and both appear to be necessary to 
its practice. But even if we could divide the methodological decisions of science into rational 
and the social, we would be left with the problem of how real some phenomena actually are, because empirical 
regularities about prejudices or tolerance are hard to be derived from entropy or temperature. 
According to recent formulations in theory of science, an advance can be achieved if the new 
theory can answer or evade more questions on an investigated subject, than the old one \cite{gro}, 
unless we maintain Thomas Kuhn position, that progress comes only through revolutions \cite{h}.
Then, the sociophysicist should be able to solve problems on social status with the second law
of thermodynamics. This task is still before us. Both for the sociology and the sociophysics, 
the subject is the same - the society. Sometimes even the empiricistic paradigm is the same. 
The barrier between these sciences comes not only from the vocabulary and the method applied, but also from 
different aspects of the social subject which are of interest for the researchers.\\

To conclude, it seems that in the near future the sociophysics will not unify with the sociology,
but rather it will develop in parallel. The area when the overlap if these sciences is most close
seems to be an analysis of large sets of data where the statistical tools work well. In such analysis,
the description of a single object is necessarily reduced to a simple characteristics. This approach
is well in spirit of the statistical physics. On the other hand, in recent decades physics has become 
concerned with objects that cannot be directly experienced by the senses. These phenomena can only be 
known through the means by which they are recorded or reasoned from indirect evidence. Indeed, it might 
be argued that their very nature is the product of theoretical description. However scientific theories 
are not simply social products with a discrete existence, but social products with complex and possibly 
diverse origins. It seems that the nature of the barriers between the sciences is mainly historical. 
If the physics itself appears to be not less subjective than the social sciences, why do not apply it to 
the society?

\bigskip


\end{document}